\documentclass[manuscript]{emulateapj}
\usepackage{longtable,amsmath}

\shorttitle{Kinematics of M31 dSphs}
\shortauthors{Collins et al.}
\def\ltsima{$\; \buildrel < \over \sim \;$}
\def\lta{\lower.5ex\hbox{\ltsima}}
\def\gtsima{$\; \buildrel > \over \sim \;$}
\def\simgt{\lower.5ex\hbox{\gtsima}}
\def\kms{{\rm\,km\,s^{-1}}}

\def\kpc{{\rm\,kpc}}

\def\msun{{\rm\,M_\odot}}
\def\lsun{{\rm\,L_\odot}}

\def\AA{$\; \buildrel \circ \over {\mathrm A}$}

\def\s{\ifmmode \widetilde \else \~\fi}
\def\={\overline}

\def\spose#1{\hbox to 0pt{#1\hss}}

\def\lta{\mathrel{\spose{\lower 3pt\hbox{$\mathchar"218$}}
     \raise 2.0pt\hbox{$\mathchar"13C$}}}
\def\gta{\mathrel{\spose{\lower 3pt\hbox{$\mathchar"218$}}
     \raise 2.0pt\hbox{$\mathchar"13E$}}}
\def\Dt{\spose{\raise 1.5ex\hbox{\hskip3pt$\mathchar"201$}}}    
\def\dt{\spose{\raise 1.0ex\hbox{\hskip2pt$\mathchar"201$}}}    

\def\sun{\odot}
\def\earth{\oplus}

\def\feh{{\rm[Fe/H]}}

\shorttitle{Kinematics in the plane - And XVI, XVII}
\shortauthors{Collins et al.}

\begin{document}
\title{Comparing the observable properties of dwarf galaxies on and off the Andromeda plane}

\author{Michelle L. M. Collins\altaffilmark{1,2,3}, Nicolas F. Martin\altaffilmark{4,1}, R. M. Rich\altaffilmark{5}, Rodrigo
  A. Ibata\altaffilmark{4}, Scott
  C. Chapman\altaffilmark{6}, Alan W. McConnachie\altaffilmark{7}, Annette M. Ferguson\altaffilmark{8}, Michael J. Irwin\altaffilmark{9}, Geraint F. Lewis\altaffilmark{10}}
\altaffiltext{1}{Max-Planck-Institut
  f\"ur Astronomie, K\"onigstuhl 17, D-69117 Heidelberg, Germany}
\altaffiltext{2} {Astronomy Department, Yale University, New Haven, CT 06510, USA }
\altaffiltext{3} {Hubble Fellow}
\altaffiltext{4}{Observatoire de Strasbourg,11, rue de l'Universit\'e,
  F-67000, Strasbourg, France}
\altaffiltext{5}{Department of Physics and Astronomy, University of
  California, Los Angeles, CA 90095-1547}
\altaffiltext{6}{Dalhousie University Dept. of Physics and Atmospheric Science
  Coburg Road Halifax, B3H1A6, Canada}
\altaffiltext{7}{NRC Herzberg Institute of Astrophysics, 5071 West Saanich
  Road, British Columbia, Victoria V9E 2E7, Canada}
\altaffiltext{8}{Institute for Astronomy, University of Edinburgh, Royal
  Observatory, Blackford Hill, Edinburgh, EH9 3HJ, UK}
\altaffiltext{9}{Institute of
 Astronomy, Madingley Rise, Cambridge, CB3 0HA ,UK}
\altaffiltext{10}{Sydney Institute
  for Astronomy, School of Physics, A28, University of Sydney, NSW 2006,
  Australia}

\begin{abstract}
  The thin, extended planes of satellite galaxies detected around
  both the Milky Way and Andromeda are not a natural prediction of the
  $\Lambda$CDM paradigm. Galaxies in these distinct planes may have formed and evolved in a different way (e.g., tidally) to their off-plane neighbours. If this were the case, one would expect the on- and
  off-plane dwarf galaxies in Andromeda to have experienced different
  evolutionary histories, which should be reflected by the chemistries,
  dynamics, and star formation histories of the two populations. In this work,
  we present new, robust kinematic observations for 2 on-plane M31 dSphs (And
  XVI and XVII) and compile and compare all available observational metrics
  for the on- and off-plane dwarfs to search for a signal that would
  corroborate such a hypothesis. We find that, barring their spatial alignment,
  the on- and off-plane Andromeda dwarf galaxies are indistinguishable from
  one another, arguing against vastly different formative and evolutionary
  histories for these two populations. 
\end{abstract}


\section{Introduction}

Dwarf spheroidal galaxies (dSphs) offer an insight into how the faintest
galaxies in the Universe have evolved over the course of cosmic
time. Those bright enough to permit the detailed study of their resolvd stellar populations are relatively nearby (distances from
$\sim20-1000\kpc$), and allow us to learn much about both their luminous
structure and their dark matter halos.

Such studies have shown that these galaxies follow
trends observed in more massive systems. Work by e.g., \citet{tolstoy09} and \citet{brasseur11b}
demonstrated that Local Group (LG) dSphs follow a well-defined
size-luminosity relationship that matches onto that of more massive late-type galaxies. They
also obey a mass-metallicity relation \citep{kirby11,kirby13a} that ties on
smoothly to that followed by dwarf irregular galaxies. Additionally, dSphs
follow a well-defined size-velocity dispersion relation \citep{collins14} that
appears to be an extrapolation of the mean rotation curve of spiral galaxies,
indicative of a mass-radius relation that holds over many orders of magnitude
in mass \citep{walker10}. 

Such relations might encourage us to think that these systems are simple to
understand. But a number of unexpected results have been unearthed in these
studies also. One particularly surprising result is the spatial alignment of
dwarf galaxies around the Milky Way (MW) and Andromeda (M31). The majority of known MW
dSphs appear to delineate a vast (diameter$\sim300\kpc$), thin (rms scale
height$\sim40\kpc$) plane structure, with a polar orientation with
respect to the MW disk (e.g.,\citealt{lyndenbell76,pawlowski12,pawlowski13a}). Similarly,
studies of M31 using data from the Pan-Andromeda Archaeological Survey (PAndAS) have revealed that $\sim40\%$
of its dwarf galaxies are aligned in a vast (diameter$\sim400\kpc$), thin (rms
scale height$\sim14\kpc$), rotating plane \citep{ibata13,conn13}. Since this work, 3 more M31 dSphs have been discovered using PanSTARRS \citep{martin13a,martin13c}, one of which (Cas III) also appears to lie in the plane, but is counter-rotating \citep{martin14b}. Such highly
ordered substructure is not a strong prediction from
cosmological simulations, where the probability of finding such thin, extended
planes is $\sim10^{-4}$ (\citealt{ibata14a,pawlowski14a}, Millennium II
simulations). Also, recent work suggests that
planes, or ordered motion of satellites, may be extremely common in both the
Local Universe, and out to $z\sim0.2$
\citep{pawlowski13a,bellazzini13,ibata14b}.  As a result, concern has been
raised as to whether the $\Lambda$-cold dark matter ($\Lambda$CDM)
 paradigm can reproduce the structured nature of substructure. In
response, alternative scenarios have been put forth to explain these planar
structures, without resting on a CDM foundation.

One such mechanism arose from
\citet{hammer10}, where the morphology of the M31 system
is explained as the result of an ancient, gas-rich merger. Follow up work by
\citet{hammer13} demonstrated that this simulation naturally reproduces a disk
of tidally created dwarf galaxies along the orbit of the merger,
morphologically similar to that observed by \citet{ibata13}. These dwarf
galaxies were created during the merger (5--8 Gyr ago) in the gas
rich tidal tails. As a result, they would not
be dark matter dominated, and would have formed in a very
different manner to dwarf galaxies off the plane, which should have
formed within their own dark matter halos. 

Such a scenario should be testable purely by comparing the properties of dwarf
galaxies both on and off the plane. Notwithstanding the question of
survivability of such tidal dwarf galaxies for Gyrs, the vastly different
formative histories should lead to differences in the kinematic, chemical and
star-forming properties of the 2 different populations. Such a comparison is
only possible in M31 as, so far, the vast majority of the identified MW dSphs
have been associated with the plane of satellites, leaving few
(cf. Sagittarius) off-plane objects for comparison.

In this {\it Letter}, we compare observational properties (sizes,
luminosities, masses, metallicities and star formation histories) of dSphs in
the M31 system. In order for the comparison to be robust, we fold in new
kinematic data for 2 on-plane dSphs, And XVI and XVII, allowing more secure
derivations of their properties. In \S2 we
present our observations and analysis of And XVI and XVII; in \S3, we compare
the properties of dSphs on- and off- the Andromeda plane of satellites and we
summarise our findings in \S4.

\section{New observations for And XVI and XVII}

Observations of And XVI and XVII were made between 1-2 October, 2013, using
the DEep Imaging Multi-Object Spectrograph (DEIMOS) on the Keck II telescope. DEIMOS is a slit-based spectrograph, and separate masks were
designed for both objects that targeted stars on the red giant
branches (RGBs) of the dSphs. To achieve spectra of the required $S/N$ for
determining velocities ($>3$\AA$^{-1}$), only stars with apparent $i$-band
magnitudes between $20.5<i<23.5$ were selected. The instrumental set-up for each mask used the 1200
line mm$^{-1}$ grating (resolution of 1.4\AA\ FWHM), and a central wavelength
of 7800\AA., giving spectral coverage from $\sim5600-9800$\AA, isolating the
region of the calcium triplet (Ca {\sc II}) at $\lambda\sim8500$\AA. The
average seeing per mask was $0.8^{\prime\prime}$ and $1.0^{\prime\prime}$ respectively.

In this work, we combine our 2013 data for And XVII with an earlier mask,
observed in September 2011 (\citealt{collins13}).  The
exposure time for the mask observed in 2011 was 3600s, while the two 2013 masks were observed for 7200s.

We reduce the resulting science spectra using a custom built pipeline, 
described in \citet{ibata11} and \citet{collins13}. We derive velocities using the Ca {\sc II} triplet absorption feature. Typically
velocity uncertainties are 3-10 kms$^{-1}$. We correct
these velocities to the heliocentric frame and for systematic shifts
caused by mis-alignments of the slits. Additionally,
for the And XVII data, we use velocities of 13 duplicate stars observed in
both masks to check the measured offsets, resulting in more
accurate velocity corrections.

\subsection{Kinematics}

\begin{figure*}
  \begin{center}
     \includegraphics[angle=0,width=0.45\hsize]{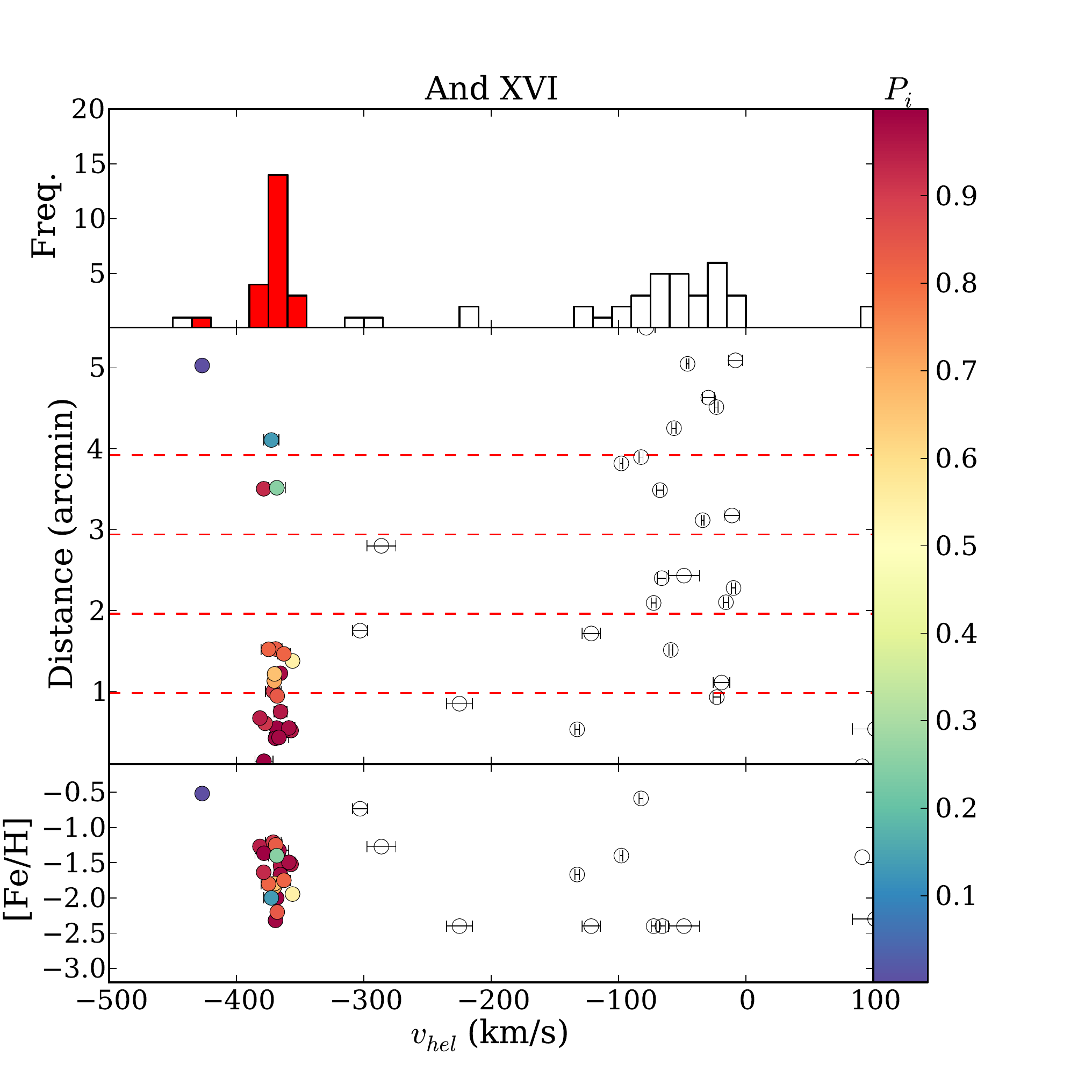}
     \includegraphics[angle=0,width=0.45\hsize]{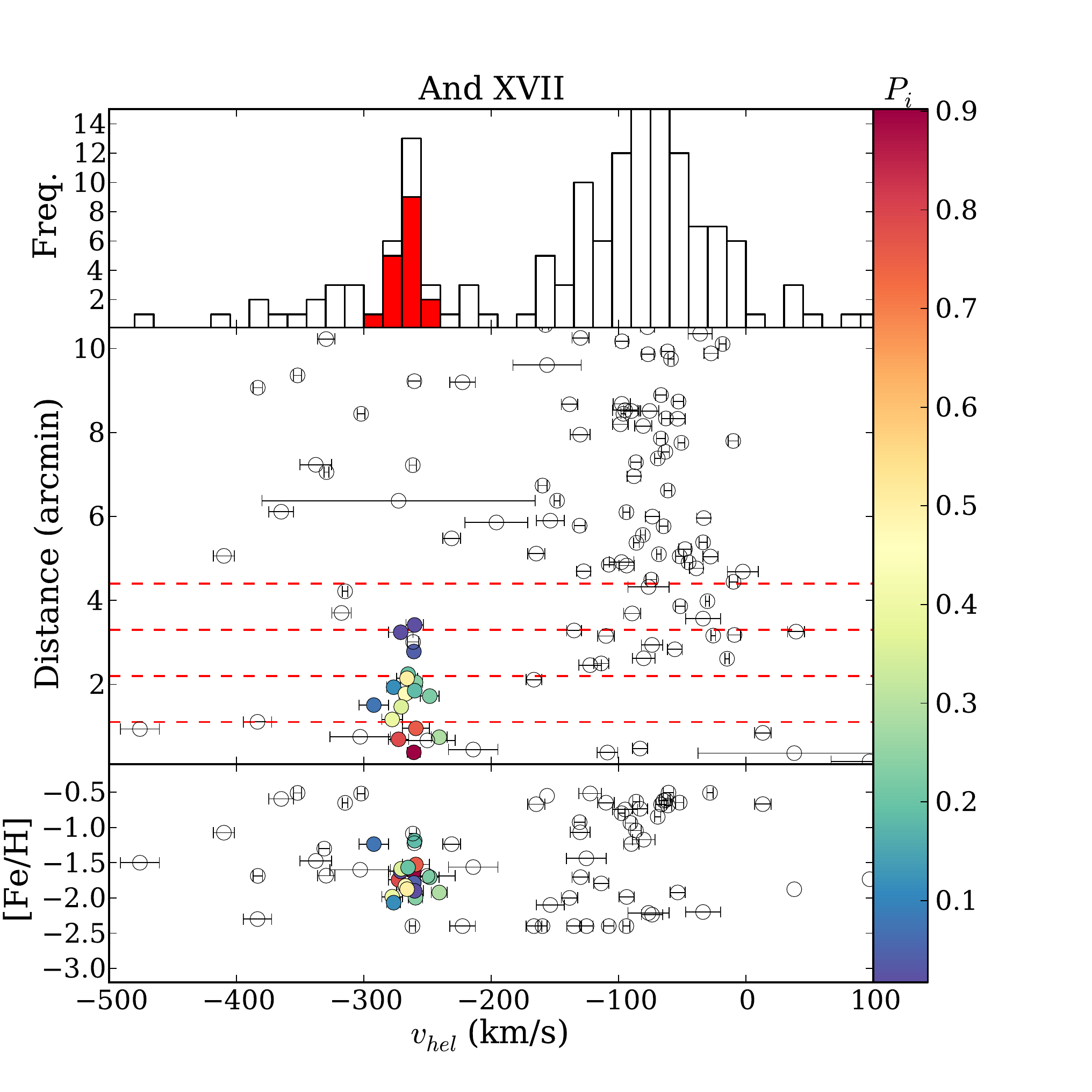}
    \includegraphics[angle=0,width=0.45\hsize]{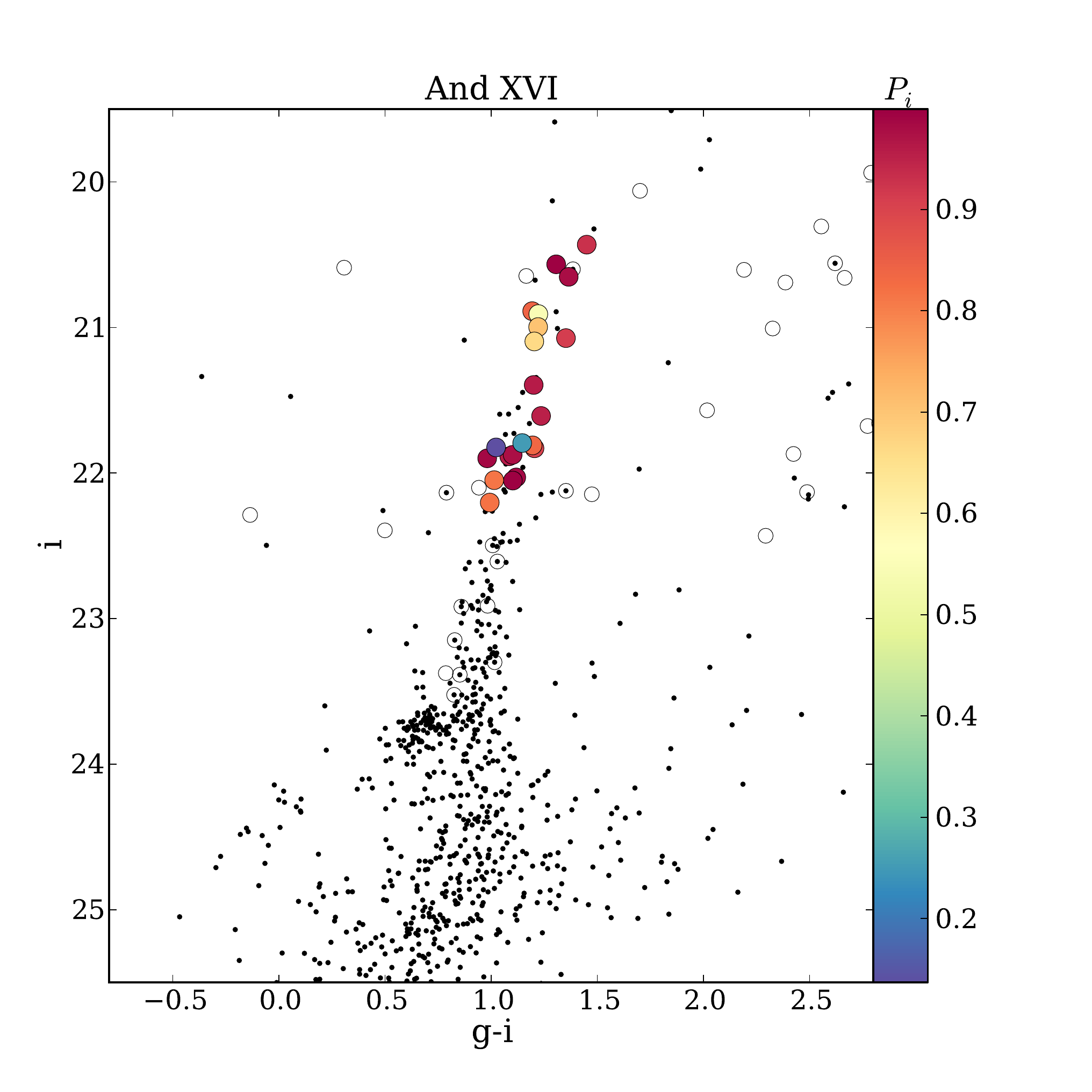}
    \includegraphics[angle=0,width=0.45\hsize]{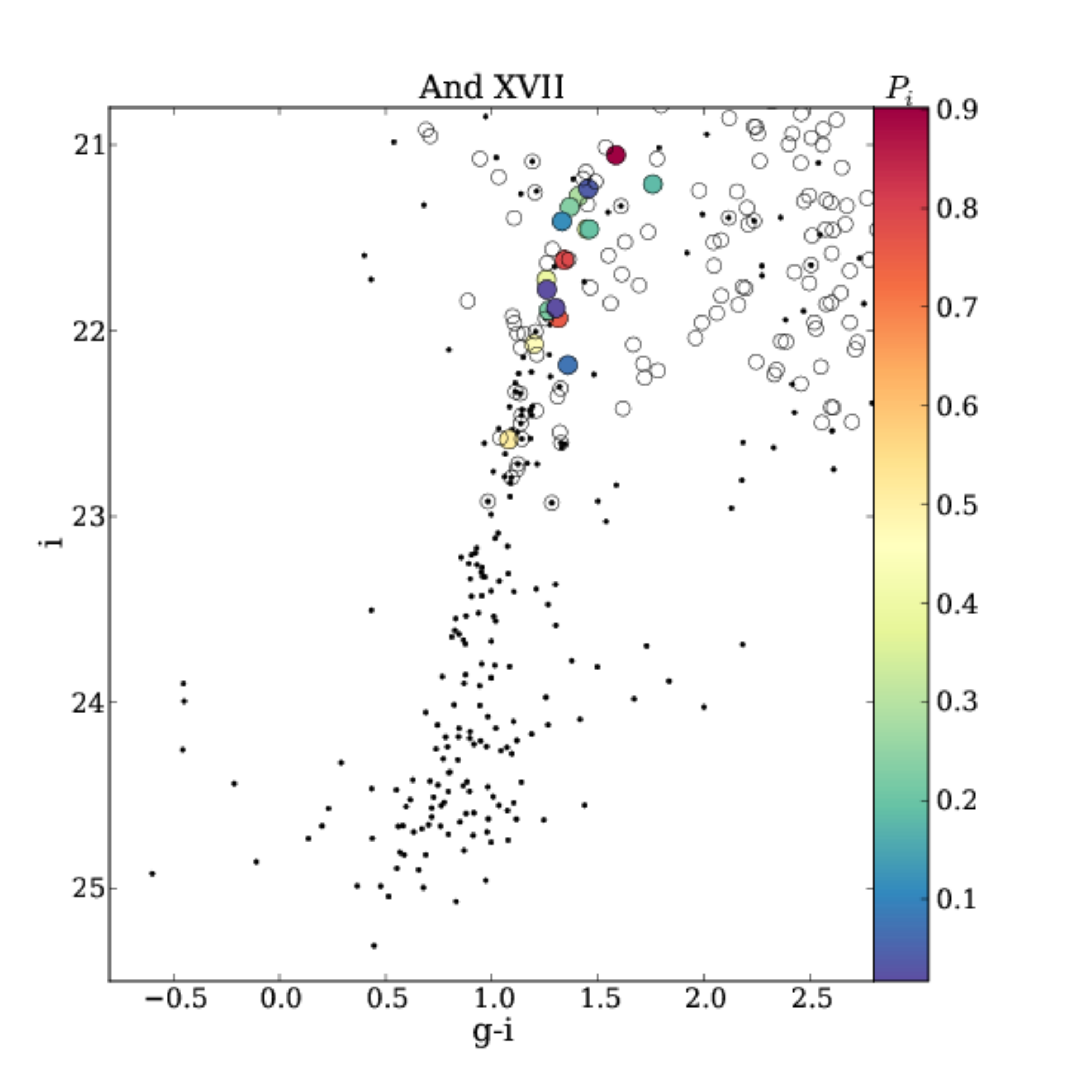}
    \caption{{\bf Top:} Kinematics for And XVI (left) and XVII (right). The
      top panels show velocity histograms of all stars observed. Probable members are highlighted as the red
      histogram. The center panel shows the distance of each star from the
      center of the dSph as a function of velocity. Here, stars are colour
      coded by their probability of membership. Stars with a negligible
      probability of membership are shown as open circles. Dashed red lines
      represent $1,2,3$ and $4\times r_{\rm half}$ for the dSph. The lower
      panel shows the photometric metallicities for stars as a function of
      velocity. Again, stars are color-coded by probability of
      membership. {\bf Bottom: } PAndAS CMDs for And XVI (left) and XVII
      (right) for all sources within 2$\times r_{\rm
        half}$ of the dSph. Stars observed with DEIMOS are color-coded by
      membership probability.}
  \label{fig:vels}
  \end{center}
\end{figure*}

For both And XVI and XVII, we aim to better constrain their
systemic velocities, $v_r$, and velocity dispersions, $\sigma_v$, as they were
previously measured from only a handful of stars (8 and 7 for And XVI and
XVII respectively). First, we determine
which observed stars are dSph members, and which are MW or M31 halo
contaminants using a probabilistic method developed by
\citet{collins13}. We assign probability of a given star being a
member of the dwarf galaxy using three criteria: (1) the position on the
color magnitude diagram of the dwarf, (2) the distance  from the center of the dwarf galaxy and (3) the
velocity. The probability of membership is the
product of these three criteria. For a detailed
description of this method, see \citet{collins13}.

In fig.~\ref{fig:vels}, we display the results of this membership
determination. The top two
plots of fig.~\ref{fig:vels} summarise the kinematic properties of the And XVI
and XVII fields. The top panels show a velocity histogram for all observed
stars. Our technique hones in on cold velocity peaks located at $\sim-370\kms$
and $\sim-260\kms$ for And XVI and XVII respectively. The central panels show
the distance from the center of the dSph as a function of radius. Here,
the points are color-coded by their probability of membership. Open points
represent stars for which the probability of membership is negligible. In the
lower panel, we show the photometrically derived [Fe/H] for each stars as a
function of velocity, determined using Dartmouth isochrones \citep{dotter08}
with age 12 Gyr and [$\alpha/$Fe]$=+0.2$.  In the lower two plots
of fig.~\ref{fig:vels}, we display the PAndAS CMDs \citep{mcconnachie09}
for both dSphs. These diagnostics isolate those stars belonging to And XVI and
XVII, indicating 20 and 16 probable member stars ($P_{\rm member}>0.1$)
respectively, more than doubling previous sample sizes. In the subsequent analysis, these probabilities act as weights for each star, allowing us to estimate all parameters for the satellites without having to make any subjective cuts on the data. In the case of And XVII, many of the stars have a low probability of membership ($P_{\rm member}\lta0.5$), as the systemic velocity of this object sits within $1\sigma$ of the M31 halo velocity ($v_{\rm r,halo}=300\kms,\sigma_{v, \rm halo}\sim90\kms$, e.g., \citealt{chapman06,kalirai06}), which is clearly visible as a non-negligable contaminant in the velocity histogram of And XVII. As such, the stars in And XVII also have non-negligible probablities of being halo contaminants. As the weights are treated relative to those of the other stars in the mask, this does not have a huge impact on measurements of the systemic velocity and dispersion, aside from increasing the uncertainties in the measurements.

Using this information, we derive $v_r$ and $\sigma_v$ for each dSph
using the grid-based maximum likelihood approach of \citet{collins13}. We
determine $v_r=-369.1^{+1.1}_{-1.3}\kms$ and $\sigma_v=5.8^{+1.1}_{-0.9}\kms$
for And XVI, and $v_r=-264.3\pm2.5\kms$ and $\sigma_v=6.5^{+3.3}_{-2.7}\kms$
for And XVII. We find that
the systemic velocity of And XVI is in good agreement with the
\citet{tollerud12} value of $v_r=-367.3\pm2.8$. The velocity dispersion we measure here is consistent with
the \citet{tollerud12} value of $\sigma_v=3.8\pm2.9\kms$, but is nominally
higher. For And XVII, we measure a significantly different $v_r$ from the
\citet{collins13} value of $v_r=-251.6^{+1.8}_{-2.0}\kms$ (almost 3$\sigma$
discrepant). This is due to our improvement in calibrating systematics in our
velocity measurements by using repeat observations
of 13 stars that are common to both masks. When stars are miscentered within their milled slits,  shifts in velocity of $10-15\kms$ can occur, and normally this is corrected for by measuring the telluric lines also imprinted onto a stars spectrum, and cross-correlating these telluric features with a rest-frame template. For faint stars (with low $S/N$) this technique can  introduce more noise into velocity measurements (as demonstrated in \citealt{collins10,collins13}). With the higher $S/N$ sources in our 2nd And XVII mask (which had double the exposure time), we were better able to correct for misalignment in the second mask, then use the velocities for the 13 duplicates to refine our velocity measurements for the 2011 mask. 

\subsection{Metallicities}

\begin{figure*}
  \begin{center}
     \includegraphics[angle=0,width=0.45\hsize]{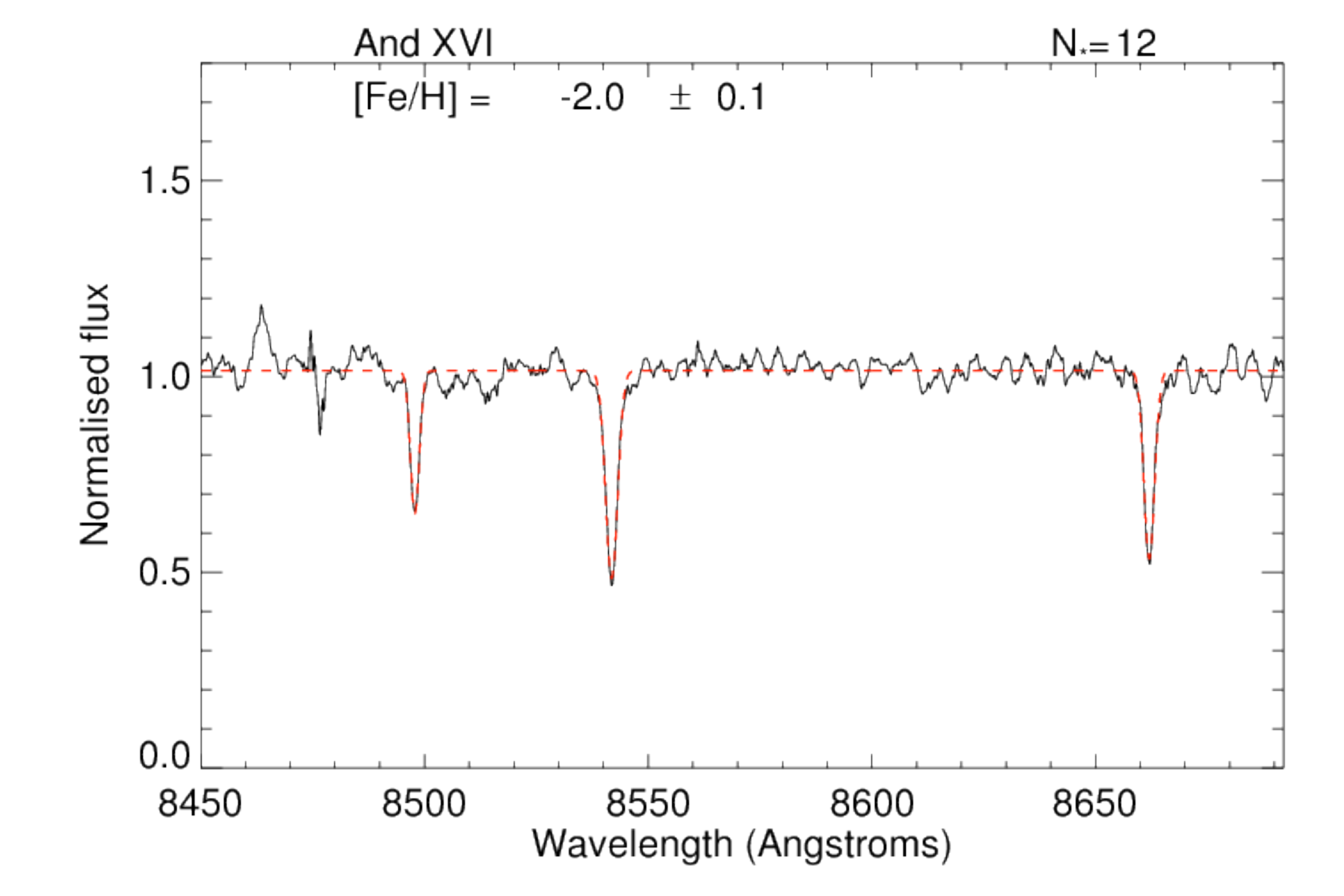}
     \includegraphics[angle=0,width=0.45\hsize]{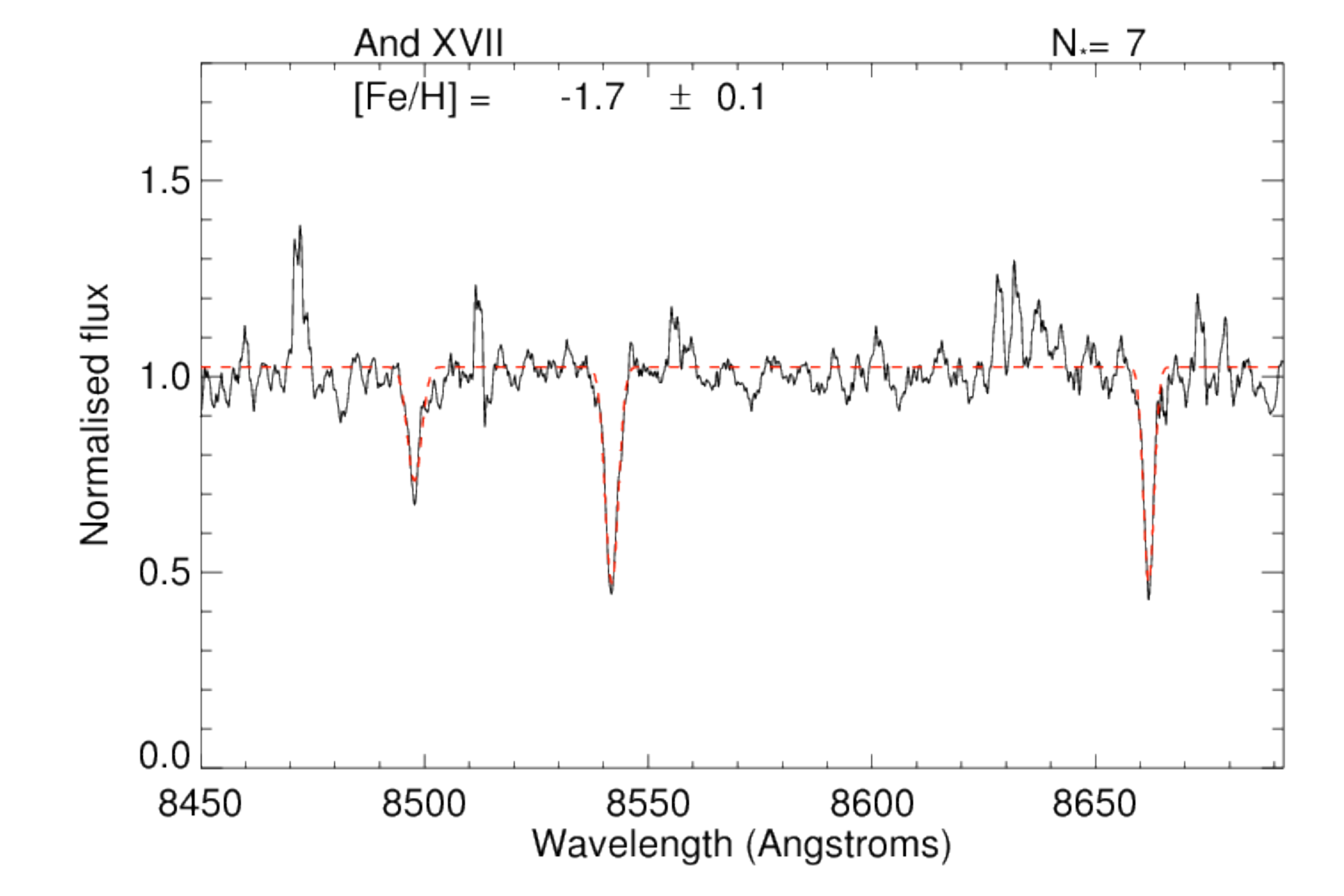}
\caption{Co-added spectra for And XVI (left) and XVII (right), constructed for
all probable member stars with $S/N>3$\AA$^{-1}$.}
  \label{fig:specfeh}
  \end{center}
\end{figure*}

We determine the average metallicities of the systems from
a co-addition of all member spectra with $S/N>3$\AA$^{-1}$ in continuum. This leaves us with a
sample of 12 stars in And XVI and 7 in And XVII. We perform a weighted
co-addition of these spectra (using both $S/N$
and $P_{\rm member}$). To determine [Fe/H], we fit the continuum and Ca {\sc II}
lines simultaneously as a polynomial and triple Gaussian. We check that each line is
uncontaminated by skylines, and then measure [Fe/H] using the \citet{starkenburg10} relation,
adapted to utilize all 3 lines of the triplet for And XVI and XVII
\citep{collins13}. The resulting spectra are shown in
fig.~\ref{fig:specfeh}. We determine $\feh=-2.0\pm0.1$ for And XVI and
$\feh=-1.7\pm0.1$ for And XVII. These values agree well with previous
spectroscopic measurements of $\feh=-2.1\pm0.2$ and $\feh=-1.9\pm0.2$ \citep{letarte09,collins13}, and photometric estimates of $\feh=-1.7$ \citep{ibata07} and $\feh=-1.9$ \cite{irwin08}.

\section{Comparing dSphs in and out of the satellite plane}

With secure kinematics for And XVI and XVII,  we possess reliable
measurements for 12 of the 14 dSphs in the M31 satellite plane. Two objects (And XI and XII)  have barely resolved velocity dispersions, so we remove them from this analysis. Using this sample, we make a global comparison of dSphs on and off the M31 satellite
plane, and search for evidence of radically different formative or
evolutionary histories for these two populations.

\begin{figure*}
  \begin{center}
     \includegraphics[angle=0,width=0.45\hsize]{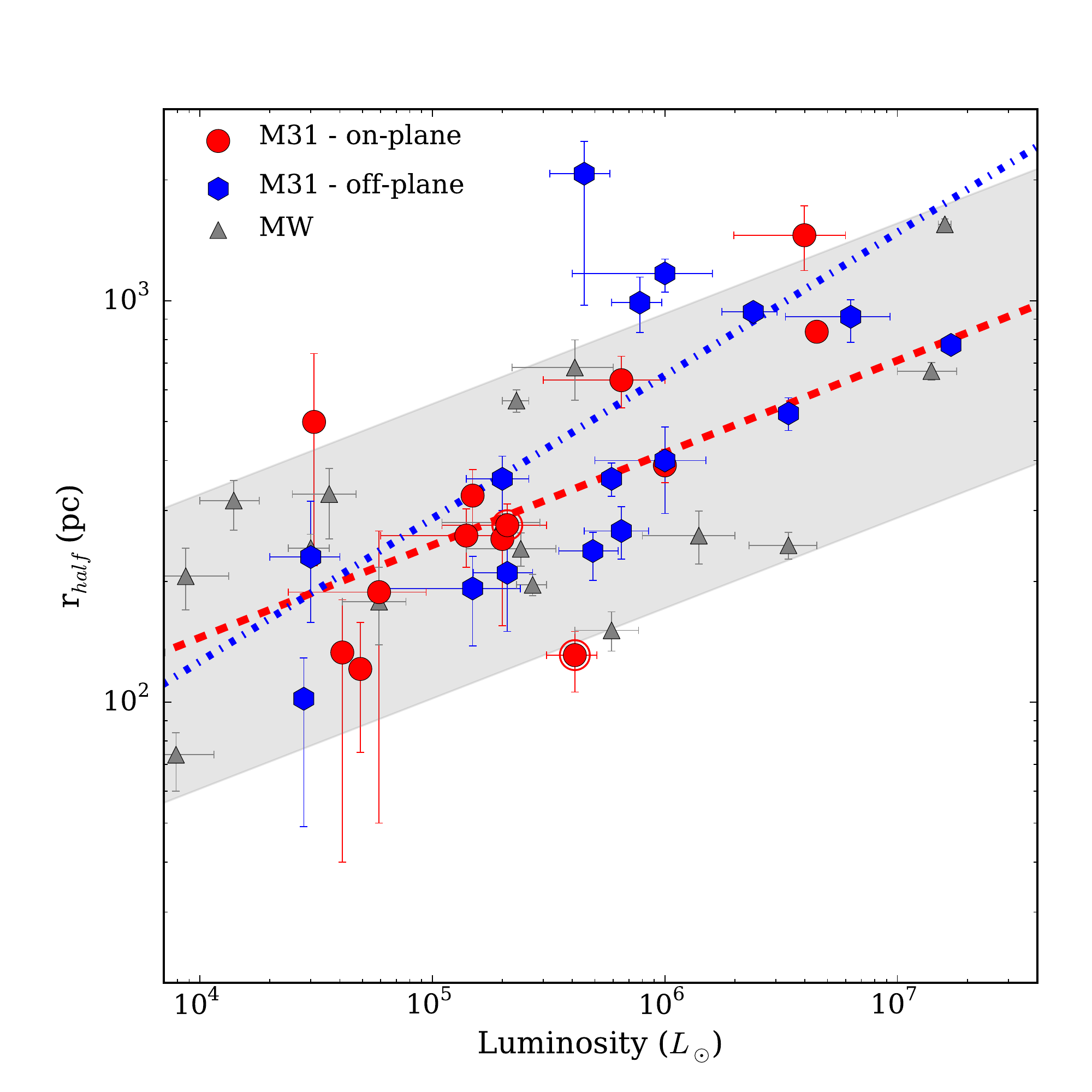}
     \includegraphics[angle=0,width=0.45\hsize]{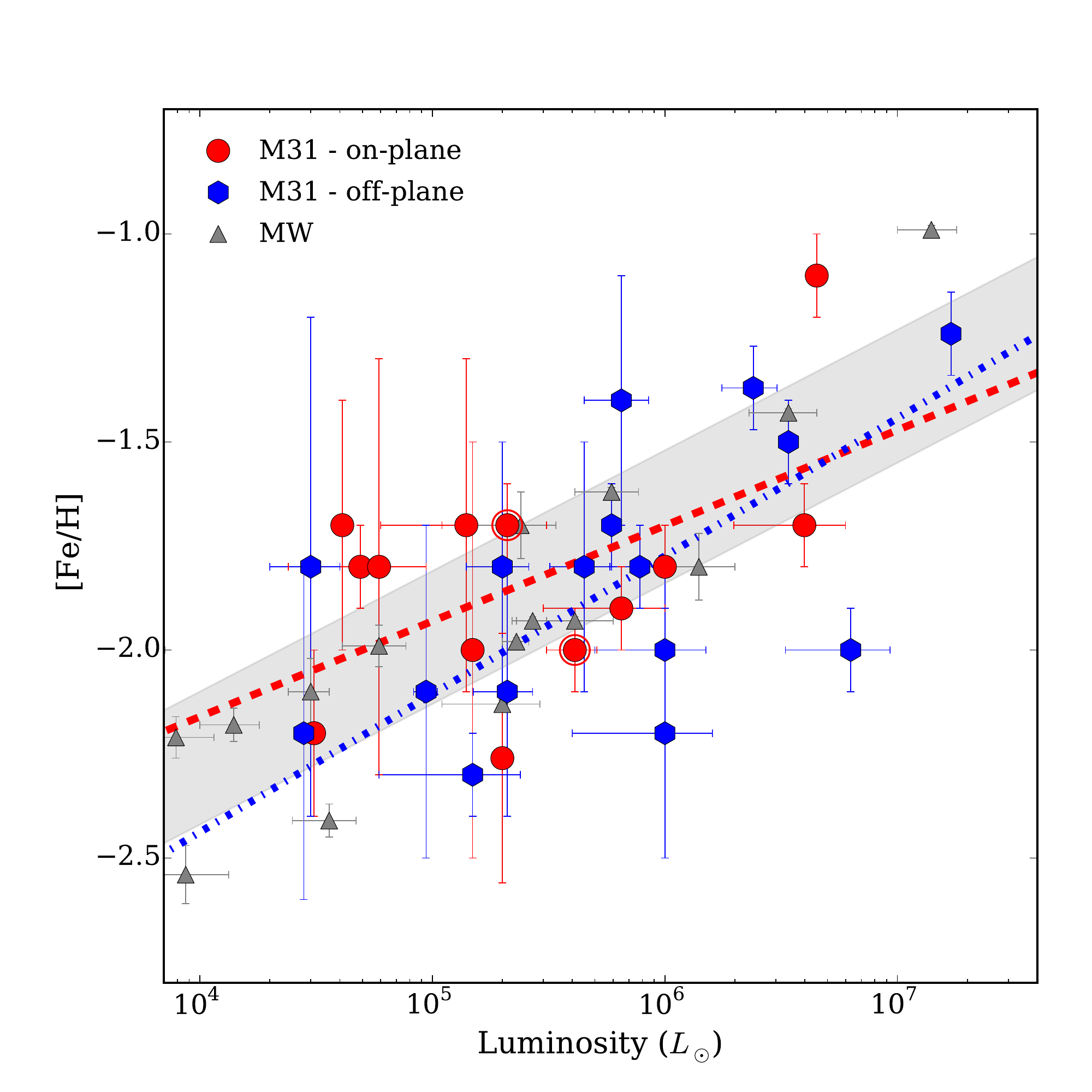}
      \includegraphics[angle=0,width=0.45\hsize]{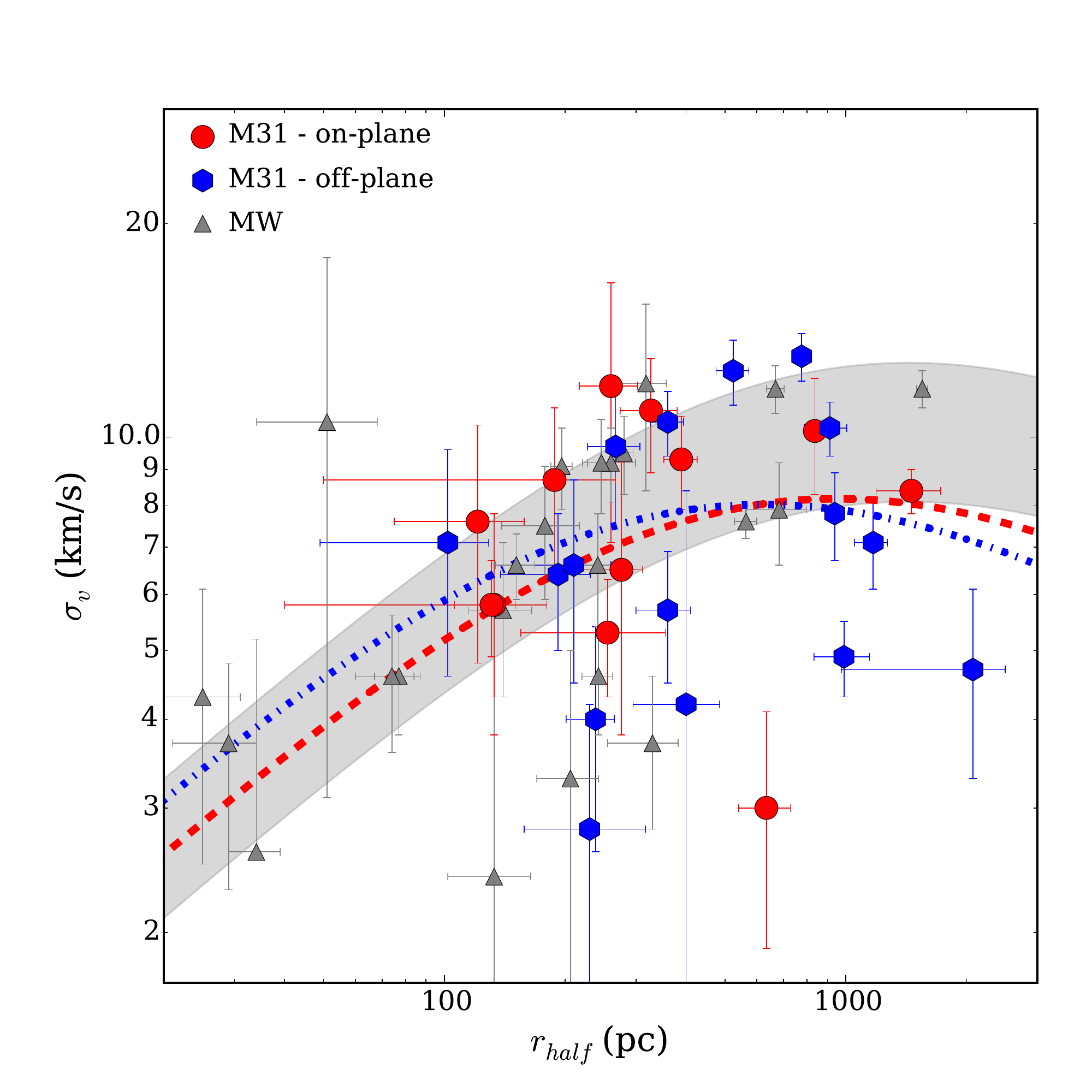}
      \includegraphics[angle=0,width=0.45\hsize]{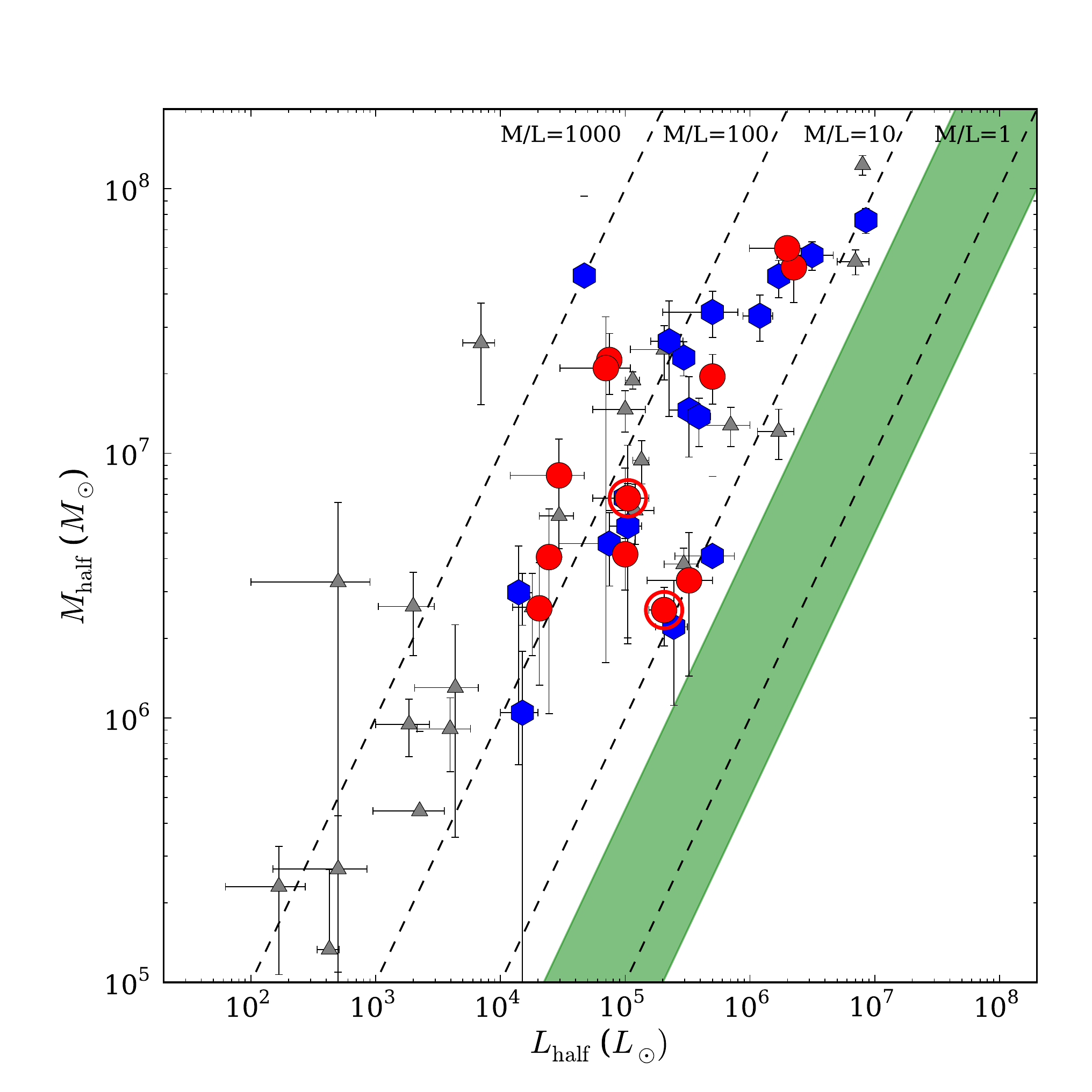}

      \caption{A comparison of the properties of
        on- and off-plane M31 satellites (shown as red circles and blue
        hexagons respectively) with MW dSphs (gray triangles), and various
        dSph relations. And XVI and XVII are shown as encircled red points. In
        all cases there are no significant differences between M31 on-plane
        and off-plane dSphs. {\bf Top left:} $L$ vs. $r_{\rm half}$ for LG
        dSphs. The \citet{brasseur11b} relation for LG dSphs is shown as the
        gray band. {\bf Top right:} $L$ vs [Fe/H] for LG dSphs. The gray band represents the \citet{kirby13a} mass-metallicity
        relation. {\bf Bottom left:} $r_{\rm half}$ vs. $\sigma_v$ for LG
        dSphs. Here, the gray band represents the range of
        NFW halo mass profiles that best encapsulate the dSphs of
        the MW and M31. {\bf Bottom right: } $L_{\rm half}$ vs. $M_{\rm half}$
        for LG dSphs. The dashed lines represent constant mass-to-light ratios
        of $1,10,100$ and $1000$ from right to left. The shaded green
        region indicates the parameter space occupied by objects with no
        significant dark matter component. All the LG dSphs fall above this
        limit, and are likely dark matter dominated.}
  \label{fig:comp}
  \end{center}
\end{figure*}

In fig.~\ref{fig:comp} we compare the different
parameter spaces probed by dynamic and photometric observations of LG dSphs. The luminosities and half-light radii for the dSphs are
collated from \citet{tolstoy09,mcconnachie12,martin13a,martin13c},  except for those M31
dSphs covered by the PAndAS survey (24 objects), where revised
measurements from Martin et al. (in prep.) are employed. The differences between the Martin et al. measurements, and previously reported values, are within $1\sigma$ of one another. The dynamics are assembled from
\citet{walker09b,koposov11,tollerud12,ho12,collins13,tollerud13,kirby13b};
Martin et al. (2014) and, for And XXI, Collins et al. (in prep). Spectroscopic
metallicities are also taken from these works, plus \citet{ho14} and
\citet{vargas14}.

The top-left panel of fig.~\ref{fig:comp} shows size vs. luminosity for LG dSph
galaxies. The gray shaded region represents the best fit
relation and $1\sigma$ scatter measured by \citet{brasseur11b} to these properties for MW and M31 dwarf
galaxies. And
XVI and XVII agree well with this relation.  To determine whether there is a significant difference between the on- and off-plane dwarfs, we perform a linear fit, where $\log(r_{\rm half})=A+B\log(M_V+6)$ to $10,000$ Monte Carlo resamples of the M31 data. The median results are plotted as the dashed and dot dashed lines in Fig.~\ref{fig:comp}. Comparing the distributions of the slope/intercepts (shown in table~\ref{tab1}), we find the on- and off-plane values agree at $<1\sigma$, meaning that the two populations appear to be
indistinguishable. We perform a similar analysis in luminosity-metallicity space (shown in the top-right panel of Fig.~\ref{fig:comp}. Here, the gray band represents the universal mass-metallicity relation of \citet{kirby13a}. We find that the linear fits to the resampled on- and off-plane data (where $\feh=A+B\log(L_V/10^6)$) are again in accord with one another at $<1\sigma$ (see table~\ref{tab1}). If the planar
dSphs had formed out of the gas rich tidal tails of a merger 5-8 Gyrs ago,
their initial chemical enrichment may have been markedly different to those
outside the plane, resutling in a different $L-\feh$ relation (e.g., \citealt{weilbacher03}). And yet there is no evidence for this in the data.

Finally, the lower 2 panels of fig.~\ref{fig:comp} inform us about the masses
and dark matter content of the MW and M31 dSphs. The left panel shows the
relationship between the size and the velocity dispersions (an indicator of
mass) of LG dSphs. The shaded region shows the range of NFW halo profiles that best represent the masses of MW
and M31 dSphs \citep{collins14}. And XVI and XVII now agree very well with
these relations, whereas previously they were tentatively low mass outliers
\citep{tollerud12,collins13}. We fit NFW profiles (\citealt{navarro97}, with the maximum circular velocity ($V_{\rm max}$) and scale radius ($R_S$) of the halo as free parameters) to our resampled on- and off-plane data. Once again, the on-plane and off-plane fits agree at $<1\sigma$  (see table~\ref{tab1}), suggesting no significant differences between these two populations. This is further
reflected in the lower right panel of fig.~\ref{fig:comp} where we present
mass (calculated from the velocity dispersion using the \citealt{walker09b} mass estimator) vs. luminosity within the half-light radius for LG dSphs. Objects that possess no significant dark
matter component (as in tidally formed dwarf galaxies) are expected to reside within the green shaded region . All the dSphs in this study are consistent with having
$[M/L]_{\rm half}\gta10\msun/\lsun$, implying that they are dark matter
dominated systems, with no apparent difference between the dSphs in the plane vs. those outside of the plane.

\begin{deluxetable}{lcc}
\tabletypesize{\footnotesize}
\tablecolumns{3} 
\tablewidth{0pt}
\tablecaption{Best fit relations for on- and off-plane M31 dSphs in size, luminosity, metallicity and mass pararmeter spaces. }
\tablehead{
\colhead{Parameter} &    \colhead{On-plane} &   \colhead{Off-plane} \\}
\startdata
 & $L$ vs. $r_{\rm half}$ &\\
\hline
Intercept, $A$ (dex) & $2.2\pm0.2$ & $2.2\pm0.2$\\
Slope, $B$ (dex)& $-0.09\pm0.07$& $-0.14\pm0.05$\\
\hline
& $L$ vs. $\feh$ & \\
\hline
Intercept, $A$ (dex)& $-1.7\pm0.1$&$-1.8\pm0.1$\\
Slope, $B$ (dex) & $0.2\pm0.2$& $0.3\pm0.2$ \\
\hline
& $r_{\rm half}$ vs. $\sigma_v$ \\
\hline
$V_{\rm max}(\kms)$ &$12.9\pm1.0$ & $13.0\pm1.2$\\
$R_S$ (pc) &$442\pm168$ & $295\pm124$\\
\enddata
\vspace{-0.3cm}
\label{tab1}
\end{deluxetable}

Another argument against a tidal formation scenario for the planar satellites
is the ages of the stars in these systems. Photometry of these systems suggest
that they are `old', and possess many stars with ages $>2$~Gyr
\citep{martin06,martin09,mcconnachie12}, and often possess RR-lyrae stars, which are at least 10 Gyrs old (e.g., And II, \citealt{pritzl04}). Recently
\citet{weisz14a} measured and compared the SFHs of And II
(an off-plane satellite) and And XVI (on-plane) using HST imaging. They found
that both had similar, extended star formation histories, with
$50-70\%$ of their stars forming 12 to 5 Gyrs ago. They were also
both quenched 5 Gyrs ago, right around the time of the merger purported to have
created the plane by \citet{hammer13}. Thus, if the plane of satellites formed
tidally, the merger that created them would need to have occurred at very early
times ($\sim10$ Gyrs ago). As this is based on only 2 objects, a complete survey of the SFHs
of M31 dSphs is necessary to validate this.

Given that, for every observation we can make in these systems, there are no measureable
differences between on- and off-plane dSphs, it is unlikely that the two populations
formed in a radically different fashion. Their spatial
orientations are all that separates them. As such, any attempt to model the
formation of this unusually thin plane must also explain the commonalities
between the on-plane and off-plane galaxies.

\section{Conclusions}

In order to create a more uniform sample for analysis of the in-plane M31 dSphs, we have presented robust kinematic properties for the
M31 dSphs, And XVI and XVII. From samples of 20 and 16 member stars
respectively, we derive
$v_r=-369.1^{+1.1}_{-1.3}\kms$ and $\sigma_v=5.8^{+1.1}_{-0.9}\kms$ for And
XVI and $v_r=-264.3\pm2.5\kms$ and $\sigma_v=6.5^{+3.3}_{-2.7}\kms$ for And
XVII. We measure average spectroscopic metallicities for both dSphs,
finding $\feh=-2.0\pm0.1$ for And XVI and $\feh=-1.7\pm0.1$ for And XVII. When
comparing their properties to those of other LG dSphs,
we find they are consistent with established trends between size, luminosity,
chemistry, and mass.

We also compare the structural and kinematic properties of 12 on-plane M31
dSphs with 18 off-plane dSphs to assess whether these two populations differ
in any way. We find that the only observation that separates them
is their spatial alignment. When comparing their sizes, luminosities, masses,
metallicities and star formation histories, these populations are
indistinguishable from one another. This argues against any radically different
formation mechanism for the on-plane dSphs, such as the
formation of these objects in a gas rich merger 5-8 Gyrs
\citep{hammer13}. Any future efforts to understand the formation of such an
unusually thin plane of satellites must therefore also account for these
universal trends.

\section*{Acknowledgments}

M.L.M.C. acknowledges funding
from the European Research Council under the EU’s
FP 7 ERC Grant Agreement n. [321035]. Support for this work was provided by NASA through Hubble Fellowship grant \#51337 awarded by the Space Telescope Science Institute, which is operated by the Association of Universities for Research in Astronomy, Inc., for NASA, under contract NAS 5-26555.

R.I. gratefully
acknowledges support from the Agence Nationale de la Recherche though
 grant POMMME (ANR 09-BLAN-0228).

N.F.M acknowledges the CNRS for support through PICS project
PICS06183.

G.F.L.  acknowledges financial support through an
ARC Future Fellowship (FT100100268) and ARC Discovery Project (DP110100678).

The data presented herein were obtained at W.M. Keck Observatory, which is
operated as a scientific partnership among the California Institute of
Technology, the University of California and the National Aeronautics and
Space Administration. The Observatory was made possible by the generous
financial support of the W.M. Keck Foundation.


\begin{thebibliography}{44}
\expandafter\ifx\csname natexlab\endcsname\relax\def\natexlab#1{#1}\fi

\bibitem[{{Bellazzini} {et~al.}(2013){Bellazzini}, {Oosterloo}, {Fraternali},
  \& {Beccari}}]{bellazzini13}
{Bellazzini}, M., {Oosterloo}, T., {Fraternali}, F., \& {Beccari}, G. 2013,
  \aap, 559, L11

\bibitem[{{Brasseur} {et~al.}(2011){Brasseur}, {Martin}, {Macci{\`o}}, {Rix},
  \& {Kang}}]{brasseur11b}
{Brasseur}, C.~M., {Martin}, N.~F., {Macci{\`o}}, A.~V., {Rix}, H.-W., \&
  {Kang}, X. 2011, \apj, 743, 179

\bibitem[{{Chapman} {et~al.}(2006){Chapman}, {Ibata}, {Lewis}, {Ferguson},
  {Irwin}, {McConnachie}, \& {Tanvir}}]{chapman06}
{Chapman}, S.~C., {Ibata}, R., {Lewis}, G.~F., {Ferguson}, A.~M.~N., {Irwin},
  M., {McConnachie}, A., \& {Tanvir}, N. 2006, \apj, 653, 255

\bibitem[{{Collins} {et~al.}(2010){Collins}, {Chapman}, {Irwin}, {Martin},
  {Ibata}, {Zucker}, {Blain}, {Ferguson}, {Lewis}, {McConnachie}, \&
  {Pe{\~n}arrubia}}]{collins10}
{Collins}, M.~L.~M., {Chapman}, S.~C., {Irwin}, M.~J., {Martin}, N.~F.,
  {Ibata}, R.~A., {Zucker}, D.~B., {Blain}, A., {Ferguson}, A.~M.~N., {Lewis},
  G.~F., {McConnachie}, A.~W., \& {Pe{\~n}arrubia}, J. 2010, \mnras, 407, 2411

\bibitem[{{Collins} {et~al.}(2013){Collins}, {Chapman}, {Rich}, {Ibata},
  {Martin}, {Irwin}, {Bate}, {Lewis}, {Pe{\~n}arrubia}, {Arimoto}, {Casey},
  {Ferguson}, {Koch}, {McConnachie}, \& {Tanvir}}]{collins13}
{Collins}, M.~L.~M., {Chapman}, S.~C., {Rich}, R.~M., {Ibata}, R.~A., {Martin},
  N.~F., {Irwin}, M.~J., {Bate}, N.~F., {Lewis}, G.~F., {Pe{\~n}arrubia}, J.,
  {Arimoto}, N., {Casey}, C.~M., {Ferguson}, A.~M.~N., {Koch}, A.,
  {McConnachie}, A.~W., \& {Tanvir}, N. 2013, \apj, 768, 172

\bibitem[{{Collins} {et~al.}(2014){Collins}, {Chapman}, {Rich}, {Ibata},
  {Martin}, {Irwin}, {Bate}, {Lewis}, {Pe{\~n}arrubia}, {Arimoto}, {Casey},
  {Ferguson}, {Koch}, {McConnachie}, \& {Tanvir}}]{collins14}
---. 2014, \apj, 783, 7

\bibitem[{{Conn} {et~al.}(2013){Conn}, {Lewis}, {Ibata}, {Parker}, {Zucker},
  {McConnachie}, {Martin}, {Valls-Gabaud}, {Tanvir}, {Irwin}, {Ferguson}, \&
  {Chapman}}]{conn13}
{Conn}, A.~R., {Lewis}, G.~F., {Ibata}, R.~A., {Parker}, Q.~A., {Zucker},
  D.~B., {McConnachie}, A.~W., {Martin}, N.~F., {Valls-Gabaud}, D., {Tanvir},
  N., {Irwin}, M.~J., {Ferguson}, A.~M.~N., \& {Chapman}, S.~C. 2013, \apj,
  766, 120

\bibitem[{{Dotter} {et~al.}(2008){Dotter}, {Chaboyer}, {Jevremovi{\'c}},
  {Kostov}, {Baron}, \& {Ferguson}}]{dotter08}
{Dotter}, A., {Chaboyer}, B., {Jevremovi{\'c}}, D., {Kostov}, V., {Baron}, E.,
  \& {Ferguson}, J.~W. 2008, \apjs, 178, 89

\bibitem[{{Hammer} {et~al.}(2013){Hammer}, {Yang}, {Fouquet}, {Pawlowski},
  {Kroupa}, {Puech}, {Flores}, \& {Wang}}]{hammer13}
{Hammer}, F., {Yang}, Y., {Fouquet}, S., {Pawlowski}, M.~S., {Kroupa}, P.,
  {Puech}, M., {Flores}, H., \& {Wang}, J. 2013, \mnras, 431, 3543

\bibitem[{{Hammer} {et~al.}(2010){Hammer}, {Yang}, {Wang}, {Puech}, {Flores},
  \& {Fouquet}}]{hammer10}
{Hammer}, F., {Yang}, Y.~B., {Wang}, J.~L., {Puech}, M., {Flores}, H., \&
  {Fouquet}, S. 2010, \apj, 725, 542

\bibitem[{{Ho} {et~al.}(2012){Ho}, {Geha}, {Munoz}, {Guhathakurta}, {Kalirai},
  {Gilbert}, {Tollerud}, {Bullock}, {Beaton}, \& {Majewski}}]{ho12}
{Ho}, N., {Geha}, M., {Munoz}, R.~R., {Guhathakurta}, P., {Kalirai}, J.,
  {Gilbert}, K.~M., {Tollerud}, E., {Bullock}, J., {Beaton}, R.~L., \&
  {Majewski}, S.~R. 2012, \apj, 758, 124

\bibitem[{{Ho} {et~al.}(2014){Ho}, {Geha}, {Tollerud}, {Zinn}, {Guhathakurta},
  \& {Vargas}}]{ho14}
{Ho}, N., {Geha}, M., {Tollerud}, E., {Zinn}, R., {Guhathakurta}, P., \&
  {Vargas}, L. 2014, ArXiv e-prints

\bibitem[{{Ibata} {et~al.}(2014{\natexlab{a}}){Ibata}, {Ibata}, {Famaey}, \&
  {Lewis}}]{ibata14b}
{Ibata}, N.~G., {Ibata}, R.~A., {Famaey}, B., \& {Lewis}, G.~F.
  2014{\natexlab{a}}, \nat, 511, 563

\bibitem[{{Ibata} {et~al.}(2007){Ibata}, {Martin}, {Irwin}, {Chapman},
  {Ferguson}, {Lewis}, \& {McConnachie}}]{ibata07}
{Ibata}, R., {Martin}, N.~F., {Irwin}, M., {Chapman}, S., {Ferguson}, A.~M.~N.,
  {Lewis}, G.~F., \& {McConnachie}, A.~W. 2007, \apj, 671, 1591

\bibitem[{{Ibata} {et~al.}(2011){Ibata}, {Sollima}, {Nipoti}, {Bellazzini},
  {Chapman}, \& {Dalessandro}}]{ibata11}
{Ibata}, R., {Sollima}, A., {Nipoti}, C., {Bellazzini}, M., {Chapman}, S.~C.,
  \& {Dalessandro}, E. 2011, \apj, 738, 186

\bibitem[{{Ibata} {et~al.}(2014{\natexlab{b}}){Ibata}, {Ibata}, {Lewis},
  {Martin}, {Conn}, {Elahi}, {Arias}, \& {Fernando}}]{ibata14a}
{Ibata}, R.~A., {Ibata}, N.~G., {Lewis}, G.~F., {Martin}, N.~F., {Conn}, A.,
  {Elahi}, P., {Arias}, V., \& {Fernando}, N. 2014{\natexlab{b}}, \apjl, 784,
  L6

\bibitem[{{Ibata} {et~al.}(2013){Ibata}, {Lewis}, {Conn}, {Irwin},
  {McConnachie}, {Chapman}, {Collins}, {Fardal}, {Ferguson}, {Ibata}, {Mackey},
  {Martin}, {Navarro}, {Rich}, {Valls-Gabaud}, \& {Widrow}}]{ibata13}
{Ibata}, R.~A., {Lewis}, G.~F., {Conn}, A.~R., {Irwin}, M.~J., {McConnachie},
  A.~W., {Chapman}, S.~C., {Collins}, M.~L., {Fardal}, M., {Ferguson},
  A.~M.~N., {Ibata}, N.~G., {Mackey}, A.~D., {Martin}, N.~F., {Navarro}, J.,
  {Rich}, R.~M., {Valls-Gabaud}, D., \& {Widrow}, L.~M. 2013, \nat, 493, 62

\bibitem[{{Irwin} {et~al.}(2008){Irwin}, {Ferguson}, {Huxor}, {Tanvir},
  {Ibata}, \& {Lewis}}]{irwin08}
{Irwin}, M.~J., {Ferguson}, A.~M.~N., {Huxor}, A.~P., {Tanvir}, N.~R., {Ibata},
  R.~A., \& {Lewis}, G.~F. 2008, \apjl, 676, L17

\bibitem[{{Kalirai} {et~al.}(2006){Kalirai}, {Gilbert}, {Guhathakurta},
  {Majewski}, {Ostheimer}, {Rich}, {Cooper}, {Reitzel}, \&
  {Patterson}}]{kalirai06}
{Kalirai}, J.~S., {Gilbert}, K.~M., {Guhathakurta}, P., {Majewski}, S.~R.,
  {Ostheimer}, J.~C., {Rich}, R.~M., {Cooper}, M.~C., {Reitzel}, D.~B., \&
  {Patterson}, R.~J. 2006, \apj, 648, 389

\bibitem[{{Kirby} {et~al.}(2013{\natexlab{a}}){Kirby}, {Boylan-Kolchin},
  {Cohen}, {Geha}, {Bullock}, \& {Kaplinghat}}]{kirby13b}
{Kirby}, E.~N., {Boylan-Kolchin}, M., {Cohen}, J.~G., {Geha}, M., {Bullock},
  J.~S., \& {Kaplinghat}, M. 2013{\natexlab{a}}, \apj, 770, 16

\bibitem[{{Kirby} {et~al.}(2013{\natexlab{b}}){Kirby}, {Cohen}, {Guhathakurta},
  {Cheng}, {Bullock}, \& {Gallazzi}}]{kirby13a}
{Kirby}, E.~N., {Cohen}, J.~G., {Guhathakurta}, P., {Cheng}, L., {Bullock},
  J.~S., \& {Gallazzi}, A. 2013{\natexlab{b}}, \apj, 779, 102

\bibitem[{{Kirby} {et~al.}(2011){Kirby}, {Lanfranchi}, {Simon}, {Cohen}, \&
  {Guhathakurta}}]{kirby11}
{Kirby}, E.~N., {Lanfranchi}, G.~A., {Simon}, J.~D., {Cohen}, J.~G., \&
  {Guhathakurta}, P. 2011, \apj, 727, 78

\bibitem[{{Koposov} {et~al.}(2011){Koposov}, {Gilmore}, {Walker}, {Belokurov},
  {Wyn Evans}, {Fellhauer}, {Gieren}, {Geisler}, {Monaco}, {Norris}, {Okamoto},
  {Penarrubia}, {Wilkinson}, {Wyse}, \& {Zucker}}]{koposov11}
{Koposov}, S.~E., {Gilmore}, G., {Walker}, M.~G., {Belokurov}, V., {Wyn Evans},
  N., {Fellhauer}, M., {Gieren}, W., {Geisler}, D., {Monaco}, L., {Norris},
  J.~E., {Okamoto}, S., {Penarrubia}, J., {Wilkinson}, M., {Wyse}, R.~F.~G., \&
  {Zucker}, D.~B. 2011, ArXiv e-prints, 1105.4102

\bibitem[{{Letarte} {et~al.}(2009){Letarte}, {Chapman}, {Collins}, {Ibata},
  {Irwin}, {Ferguson}, {Lewis}, {Martin}, {McConnachie}, \&
  {Tanvir}}]{letarte09}
{Letarte}, B., {Chapman}, S.~C., {Collins}, M., {Ibata}, R.~A., {Irwin}, M.~J.,
  {Ferguson}, A.~M.~N., {Lewis}, G.~F., {Martin}, N., {McConnachie}, A., \&
  {Tanvir}, N. 2009, \mnras, 400, 1472

\bibitem[{{Lynden-Bell}(1976)}]{lyndenbell76}
{Lynden-Bell}, D. 1976, \mnras, 174, 695


\bibitem[{{Martin} {et~al.}(2006){Martin}, {Ibata}, {Irwin}, {Chapman},
  {Lewis}, {Ferguson}, {Tanvir}, \& {McConnachie}}]{martin06}
{Martin}, N.~F., {Ibata}, R.~A., {Irwin}, M.~J., {Chapman}, S., {Lewis}, G.~F.,
  {Ferguson}, A.~M.~N., {Tanvir}, N., \& {McConnachie}, A.~W. 2006, \mnras,
  371, 1983

\bibitem[{{Martin} {et~al.}(2009){Martin}, {McConnachie}, {Irwin}, {Widrow},
  {Ferguson}, {Ibata}, {Dubinski}, {Babul}, {Chapman}, {Fardal}, {Lewis},
  {Navarro}, \& {Rich}}]{martin09}
{Martin}, N.~F., {McConnachie}, A.~W., {Irwin}, M., {Widrow}, L.~M.,
  {Ferguson}, A.~M.~N., {Ibata}, R.~A., {Dubinski}, J., {Babul}, A., {Chapman},
  S., {Fardal}, M., {Lewis}, G.~F., {Navarro}, J., \& {Rich}, R.~M. 2009, \apj,
  705, 758

\bibitem[{{Martin} {et~al.}(2013{\natexlab{a}}){Martin}, {Schlafly}, {Slater},
  {Bernard}, {Rix}, {Bell}, {Ferguson}, {Finkbeiner}, {Laevens}, {Burgett},
  {Chambers}, {Draper}, {Hodapp}, {Kaiser}, {Kudritzki}, {Magnier}, {Metcalfe},
  {Morgan}, {Price}, {Tonry}, {Wainscoat}, \& {Waters}}]{martin13c}
{Martin}, N.~F., {Schlafly}, E.~F., {Slater}, C.~T., {Bernard}, E.~J., {Rix},
  H.-W., {Bell}, E.~F., {Ferguson}, A.~M.~N., {Finkbeiner}, D.~P., {Laevens},
  B.~P.~M., {Burgett}, W.~S., {Chambers}, K.~C., {Draper}, P.~W., {Hodapp},
  K.~W., {Kaiser}, N., {Kudritzki}, R.-P., {Magnier}, E.~A., {Metcalfe}, N.,
  {Morgan}, J.~S., {Price}, P.~A., {Tonry}, J.~L., {Wainscoat}, R.~J., \&
  {Waters}, C. 2013{\natexlab{a}}, \apjl, 779, L10

\bibitem[{{Martin} {et~al.}(2013{\natexlab{b}}){Martin}, {Slater}, {Schlafly},
  {Morganson}, {Rix}, {Bell}, {Laevens}, {Bernard}, {Ferguson}, {Finkbeiner},
  {Burgett}, {Chambers}, {Hodapp}, {Kaiser}, {Kudritzki}, {Magnier}, {Morgan},
  {Price}, {Tonry}, \& {Wainscoat}}]{martin13a}
{Martin}, N.~F., {Slater}, C.~T., {Schlafly}, E.~F., {Morganson}, E., {Rix},
  H.-W., {Bell}, E.~F., {Laevens}, B.~P.~M., {Bernard}, E.~J., {Ferguson},
  A.~M.~N., {Finkbeiner}, D.~P., {Burgett}, W.~S., {Chambers}, K.~C., {Hodapp},
  K.~W., {Kaiser}, N., {Kudritzki}, R.-P., {Magnier}, E.~A., {Morgan}, J.~S.,
  {Price}, P.~A., {Tonry}, J.~L., \& {Wainscoat}, R.~J. 2013{\natexlab{b}},
  \apj, 772, 15

\bibitem[{{Martin} {et~al.}(2014){Martin},{Chambers},{Collins}, {Ibata}, {Rich},{Bell},{Bernard},{Ferguson},{Flewelling},{Kaiser},{Magnier},  {Tonry}, {Wainscoat}}]{martin14b}
{Martin}, N.~F.  {Chambers}, K.~C. {Collins}, M.~L.~M.  
	{Ibata}, R.~A.{Rich}, R.~M.  {Bell}, E.~F., {Bernard}, E.~J., 
	{Ferguson}, A.~M.~N., {Flewelling}, H., {Kaiser}, N., 
	{Magnier}, E.~A., {Tonry}, J.~L., {Wainscoat}, R.~J.,\apjl,793,14

\bibitem[{{McConnachie}(2012)}]{mcconnachie12}
{McConnachie}, A.~W. 2012, \aj, 144, 4

\bibitem[{{McConnachie} {et~al.}(2009)}]{mcconnachie09}
{McConnachie}, A.~W. {et~al.} 2009, \nat, 461, 66

\bibitem[{{Navarro} {et~al.}(1997){Navarro}, {Frenk}, \& {White}}]{navarro97}
{Navarro}, J.~F., {Frenk}, C.~S., \& {White}, S.~D.~M. 1997, \apj, 490, 493

\bibitem[{{Pawlowski} {et~al.}(2014){Pawlowski}, {Famaey}, {Jerjen}, {Merritt},
  {Kroupa}, {Dabringhausen}, {L{\"u}ghausen}, {Forbes}, {Hensler}, {Hammer},
  {Puech}, {Fouquet}, {Flores}, \& {Yang}}]{pawlowski14a}
{Pawlowski}, M.~S., {Famaey}, B., {Jerjen}, H., {Merritt}, D., {Kroupa}, P.,
  {Dabringhausen}, J., {L{\"u}ghausen}, F., {Forbes}, D.~A., {Hensler}, G.,
  {Hammer}, F., {Puech}, M., {Fouquet}, S., {Flores}, H., \& {Yang}, Y. 2014,
  \mnras, 442, 2362

\bibitem[{{Pawlowski} {et~al.}(2013){Pawlowski}, {Kroupa}, \&
  {Jerjen}}]{pawlowski13a}
{Pawlowski}, M.~S., {Kroupa}, P., \& {Jerjen}, H. 2013, \mnras, 435, 1928

\bibitem[{{Pawlowski} {et~al.}(2012){Pawlowski}, {Pflamm-Altenburg}, \&
  {Kroupa}}]{pawlowski12}
{Pawlowski}, M.~S., {Pflamm-Altenburg}, J., \& {Kroupa}, P. 2012, \mnras, 423,
  1109

\bibitem[{{Pritzl} {et~al.}(2004){Pritzl}, {Armandroff}, {Jacoby}, \& {Da
  Costa}}]{pritzl04}
{Pritzl}, B.~J., {Armandroff}, T.~E., {Jacoby}, G.~H., \& {Da Costa}, G.~S.
  2004, \aj, 127, 318


\bibitem[{{Starkenburg} {et~al.}(2010){Starkenburg}, {Hill}, {Tolstoy},
  {Gonz{\'a}lez Hern{\'a}ndez}, {Irwin}, {Helmi}, {Battaglia}, {Jablonka},
  {Tafelmeyer}, {Shetrone}, {Venn}, \& {de Boer}}]{starkenburg10}
{Starkenburg}, E., {Hill}, V., {Tolstoy}, E., {Gonz{\'a}lez Hern{\'a}ndez},
  J.~I., {Irwin}, M., {Helmi}, A., {Battaglia}, G., {Jablonka}, P.,
  {Tafelmeyer}, M., {Shetrone}, M., {Venn}, K., \& {de Boer}, T. 2010, \aap,
  513, A34+

\bibitem[{{Tollerud} {et~al.}(2012){Tollerud}, {Beaton}, {Geha}, {Bullock},
  {Guhathakurta}, {Kalirai}, {Majewski}, {Kirby}, {Gilbert}, {Yniguez},
  {Patterson}, {Ostheimer}, {Cooke}, {Dorman}, {Choudhury}, \&
  {Cooper}}]{tollerud12}
{Tollerud}, E.~J., {Beaton}, R.~L., {Geha}, M.~C., {Bullock}, J.~S.,
  {Guhathakurta}, P., {Kalirai}, J.~S., {Majewski}, S.~R., {Kirby}, E.~N.,
  {Gilbert}, K.~M., {Yniguez}, B., {Patterson}, R.~J., {Ostheimer}, J.~C.,
  {Cooke}, J., {Dorman}, C.~E., {Choudhury}, A., \& {Cooper}, M.~C. 2012, \apj,
  752, 45

\bibitem[{{Tollerud} {et~al.}(2013){Tollerud}, {Geha}, {Vargas}, \&
  {Bullock}}]{tollerud13}
{Tollerud}, E.~J., {Geha}, M.~C., {Vargas}, L.~C., \& {Bullock}, J.~S. 2013,
  ArXiv e-prints

\bibitem[{{Tolstoy} {et~al.}(2009){Tolstoy}, {Hill}, \& {Tosi}}]{tolstoy09}
{Tolstoy}, E., {Hill}, V., \& {Tosi}, M. 2009, \araa, 47, 371

\bibitem[{{Vargas} {et~al.}(2014){Vargas}, {Geha}, \& {Tollerud}}]{vargas14}
{Vargas}, L.~C., {Geha}, M.~C., \& {Tollerud}, E.~J. 2014, \apj, 790, 73

\bibitem[{{Walker} {et~al.}(2009){Walker}, {Mateo}, {Olszewski},
  {Pe{\~n}arrubia}, {Wyn Evans}, \& {Gilmore}}]{walker09b}
{Walker}, M.~G., {Mateo}, M., {Olszewski}, E.~W., {Pe{\~n}arrubia}, J., {Wyn
  Evans}, N., \& {Gilmore}, G. 2009, \apj, 704, 1274

\bibitem[{{Walker} {et~al.}(2010){Walker}, {McGaugh}, {Mateo}, {Olszewski}, \&
  {Kuzio de Naray}}]{walker10}
{Walker}, M.~G., {McGaugh}, S.~S., {Mateo}, M., {Olszewski}, E.~W., \& {Kuzio
  de Naray}, R. 2010, \apjl, 717, L87

\bibitem[{{Weilbacher} {et~al.}(2003){Weilbacher}, {Duc}, \&
  {Fritze-v.~Alvensleben}}]{weilbacher03}
{Weilbacher}, P.~M., {Duc}, P.-A., \& {Fritze-v.~Alvensleben}, U. 2003, \aap,
  397, 545


\bibitem[{{Weisz} {et~al.}(2014){Weisz}, {Skillman}, {Hidalgo}, {Monelli},
  {Dolphin}, {McConnachie}, {Bernard}, {Gallart}, {Aparicio}, {Boylan-Kolchin},
  {Cassisi}, {Cole}, {Ferguson}, {Irwin}, {Martin}, {Mayer}, {McQuinn},
  {Navarro}, \& {Stetson}}]{weisz14a}
{Weisz}, D.~R., {Skillman}, E.~D., {Hidalgo}, S.~L., {Monelli}, M., {Dolphin},
  A.~E., {McConnachie}, A., {Bernard}, E.~J., {Gallart}, C., {Aparicio}, A.,
  {Boylan-Kolchin}, M., {Cassisi}, S., {Cole}, A.~A., {Ferguson}, H.~C.,
  {Irwin}, M., {Martin}, N.~F., {Mayer}, L., {McQuinn}, K.~B.~W., {Navarro},
  J.~F., \& {Stetson}, P.~B. 2014, \apj, 789, 24

\end{thebibliography}

\end{document}